\documentclass[conference]{IEEEtran}
\usepackage{graphicx}
\usepackage{xurl}
\usepackage[labelformat=simple]{subcaption}

\usepackage{algorithm,amsbsy,amsfonts,amssymb,amsmath,amssymb,bbm,mathrsfs,multirow,amsthm,epstopdf,cite}
\usepackage{graphicx, caption,subcaption}
\usepackage{setspace} 
\usepackage{amsmath}
\usepackage{algorithm}
\usepackage{algpseudocode}

\usepackage[table]{xcolor}
\usepackage{diagbox}
\usepackage[labelformat=simple]{subcaption}

\usepackage[top=0.75in, bottom=0.95in, left=0.67in, right=0.67in]{geometry}

\DeclareMathAlphabet\mathbfcal{OMS}{cmsy}{b}{n}
\algnewcommand{\LineComment}[1]{\State \(\triangleright\) #1}

\newtheorem{theorem}{\textbf{\text{Theorem}}}

\begin{document}
	\title{A Novel Blockchain-Based Information Management Framework for Web 3.0}
	\author{Md Arif Hassan, Cong T. Nguyen, Chi-Hieu Nguyen, Dinh Thai Hoang, Diep N. Nguyen, and Eryk Dutkiewicz  \\		
		School of Electrical and Data Engineering, University of Technology Sydney, Australia \\
		\vspace{-5mm}	}

	\maketitle
%++++++++++++++++++++++++++++++++++++++++++++++++++++++++++++++++++++
%++++++++++++++++++++++++++++++++++++++++++++++++++++++++++++++++++++
\begin{abstract}	
Web 3.0 is the third generation of the World Wide Web (WWW), concentrating on the critical concepts of decentralization, availability, and increasing client usability. Although Web 3.0 is undoubtedly an essential component of the future Internet, it currently faces critical challenges, including decentralized data collection and management. To overcome these challenges, blockchain has emerged as one of the core technologies for the future development of Web 3.0. In this paper, we propose a novel blockchain-based information management framework, namely Smart Blockchain-based Web (SBW), to manage information in Web 3.0 effectively, enhance the security and privacy of users’ data, bring additional profits, and incentivize users to contribute information to the websites. Particularly, SBW utilizes blockchain technology and smart contracts to manage the decentralized data collection process for Web 3.0 effectively. Moreover, in this framework, we develop an effective consensus mechanism based on Proof-of-Stake (PoS) to reward the user's information contribution and conduct game theoretical analysis to analyze the user’s behavior in the considered system. Additionally, we conduct simulations to assess the performance of SBW and investigate the impact of critical parameters on information contribution. The findings confirm our theoretical analysis and demonstrate that our proposed consensus mechanism can incentivize the nodes and users to contribute more information to our systems.
\end{abstract}

{\it Keywords-} Blockchain, Web 3.0, Proof-of-Stake, smart contract, and game theory.
	
\thispagestyle{empty}

%++++++++++++++++++++++++++++++++++++++++++++++++++++++++++++++++++++
%++++++++++++++++++++++++++++++++++++++++++++++++++++++++++++++++++++
\section{Introduction}

\subsection{Motivations}
Since the infancy of the World Wide Web (WWW) in the early 1990s, web technology has undergone several modifications, especially Web 1.0 and Web 2.0, which have impacted how users interact with the Internet. Web 1.0, the initial generation of the internet, was mostly made up of static HTML pages with limited interaction from users.  Web 2.0 introduced evolution towards dynamic content, increased user engagement, and the prevalence of social networking platforms. Web 3.0 has been considered the next generation of the WWW, concentrating on decentralization and user-centricity~\cite{Liu}. The development of Web 3.0 is expected to create a more intelligent and interconnected version of the web, where machines and software programs can understand and represent information in a more creative manner. This would allow more effective communication between machines and humans and more efficient search and discovery of information. To achieve this goal, Web 3.0 relies on several technologies, such as machine learning, natural language processing, artificial intelligence, and using ontologies and metadata to describe and categorize data~\cite{singlegrain}.

Compared to the conventional web, Web 3.0  offers several advantages, including decentralized data, information governance, accessibility, and uninterrupted performance~\cite{techutzpah}. However, due to the characteristics of using decentralized networks, the implementation of Web 3.0 is facing many difficulties and challenges in practice, especially related to storing and processing data in a decentralized manner~\cite{Liu}. Specifically, because data is stored and processed at many network nodes simultaneously, we need to ensure that this data is authenticated properly and presented on websites consistently without duplications and redundancies. There are also many other important issues in storing and processing decentralized data that Web 3.0 is facing, including network scalability, data privacy and security, and interoperability problems~\cite{wang}. Therefore, novel solutions are urgently needed for the future development of Web 3.0.

Blockchain technology~\cite{r1} is a distributed digital ledger that is being increasingly adopted as a key component of the evolution of Web 3.0. Using blockchain in Web 3.0 enables the creation of decentralized systems that can operate without a central authority, resulting in better security, transparency, and reliability in managing data and transactions. The development of blockchain technology for Web 3.0 requires the creation of robust and scalable protocols that can handle the large volumes of data and transactions that are expected to be processed by these systems. Additionally, there is a need to address issues related to scalability, privacy, and interoperability and ensure that the technology is user-friendly and accessible to a wide range of users. The integration of blockchain into the development of Web 3.0 has the potential to significantly transform how the Internet functions, paving the way for innovative services that are more secure, transparent, and decentralized.

\subsection{Related Works}
Thanks to its outstanding features, more blockchain-based approaches have been proposed for Web 3.0. For example, a blockchain-based Web 3.0 framework is developed in~\cite{Shawon2021} to manage educational certificates for graduate students and universities. Particularly, the issuance and verification processes for certificates are carried out using Ethereum and smart contracts without a centralized authority. Similarly, a blockchain-based Web 3.0 framework for health contract management is proposed in~\cite{Chondrogiannis2022}. In this study, the authors create a Web 3.0 framework utilizing blockchain technology that enables people and health insurance providers to negotiate contracts using health standards and semantic web technologies. Once created, these health contracts are stored in the blockchain, and thus, they benefit from the data immutability and integrity of blockchain technologies. A similar approach to patient healthcare employing blockchain and semantic web technology is introduced~\cite{delaney2021present} to address data availability and patient privacy protection. Particularly, health data is stored on the blockchain to protect patient privacy, and when clinicians query health data, semantic web technology is employed to provide relevant data to the clinicians. 

Regarding the development of blockchain technology for Web 3.0, there are several works introduced recently. For example, a Proof-of-Work (PoW) based asymmetric encryption consensus is proposed in~\cite{yang2022} for Web 3.0. In this study, the authors create a Timed-Release Encryption where information is encrypted on a blockchain using an asymmetric key encryption mechanism without relying on any external agents. The block time in a PoW blockchain is dynamically managed, so the release time is predictable. However, this study has several limitations, such as altering the difficulty based on the bit length of the prime number. Alternatively, researchers also introduce various new consensus mechanisms, such as proof-of-semantic, to develop Web 3.0. For example, a blockchain-semantic wireless edge for Web 3.0 is presented in~\cite{Lin2022}, in which the authors attempt to provide a secure proof-of-semantic framework to execute dynamic connections of Web 3.0 on semantic validation techniques. In this study, the authors built an adaptive DRL-based sharding Oracle technique to boost communication efficiency, which mitigates latencies and accomplishes a multiplicity of Web 3.0 activities. The authors also conduct a case study demonstrating how the proposed framework may dynamically alter Oracle settings based on changing semantic demands.

Previous works such as~\cite{Shawon2021}, \cite{Chondrogiannis2022},\cite{yang2022} rely on the PoW consensus mechanism that depends on a computing power competition among individuals to obtain consensus. As a result, the PoW process consumes a significant amount of energy and has a substantial latency~\cite{digiconomist}. Furthermore, a decentralized data management system must be developed to ensure that data is always available, can be updated seamlessly, and allows people to receive rewards for sharing information. To overcome those limitations, the Proof-of-Stake (PoS) consensus has been developed recently, which replaces the computing power competition with a stake ownership competition. Therefore, the PoS mechanism significantly benefits PoW, such as the lowest energy consumption and high transaction processing capabilities~\cite{nguyen2019proof}. Thanks to these advantages, compared to PoW, the PoS mechanism is more beneficial for Web 3.0 applications which are expected to handle a huge amount of data from many users.

\subsection{Contributions of the Paper}

In this paper, we develop a blockchain-based framework for Web 3.0, namely SBW 3.0, which can effectively manage information in Web 3.0, bring additional profits to users, and incentivize users to contribute information to the web services. Notably, in our framework, a smart contract is created and stored in the blockchain when a user wants to contribute information to the web. Afterward, when a user sends information to a mining node, it will automatically pay the user for their information contribution. Moreover, we develop a new consensus mechanism based on PoS that rewards the decentralized nodes for their information contribution, thereby incentivizing them to contribute information to the website. Furthermore, we conduct a game theoretical analysis to analyze our system's user behavior. The obtained result can bring additional benefits to the users as well as incentivize users to contribute more information to the website. Additionally, we perform various simulations to investigate the impact of significant factors on the information contribution. The outcomes show that the block reward can play a key role in significantly improving users' information contribution. The critical contributions of the proposed framework can be summarized as follows:
\vspace{-2pt}
\begin{itemize}
 \item We develop a novel blockchain-based framework that leverages the advantages of smart contracts to effectively manage the information exchange and payment process for Web 3.0. 
\item We propose an intelligent PoS-based consensus mechanism to reward users' information contribution, thereby incentivizing users to contribute information to the proposed SBW 3.0. Moreover, we conduct a game theoretical analysis to analyze users' behaviors in the proposed system. These analyses are very useful for system designers to deploy appropriate parameters to optimize.
 \item We conduct extensive simulations to assess the performance of our proposed framework. The results confirm our theoretical analysis and show that our proposed consensus mechanism can bring additional profits to the users and incentivize users to contribute more to the system. Furthermore, we study essential factors to examine their effects on the system performance, which can significantly improve users' information contribution.
 \end{itemize}

	\section{Blockchain-based Information Management for Web 3.0}
	\label{sec:SM}
	
Blockchain technology has proven to be a promising solution to address decentralized data management challenges. Information can be stored by multiple decentralized nodes instead of a central server. Thus, even when several nodes are down, the website still works, thanks to other active nodes. Moreover, the information processing workload can be shared among multiple nodes, and thus, the increased demands for information processing can be addressed. Another benefit of this decentralized information processing is that the information sources can send their data to nearby nodes, unlike a central server (which is physically far away). As a result, the delay in transmission can be reduced significantly. Furthermore, since no single authority controls the website, problems such as censorship and privacy can be mitigated. Therefore, in the following section, we propose a novel framework, namely Smart Blockchain-based Web 3.0 (SBW), for effective information management for Web 3.0.
\vspace{-4pt}	

\subsection{Proposed System Model}
\begin{figure}[!]
		\centering
		\includegraphics[width=1.0\linewidth]{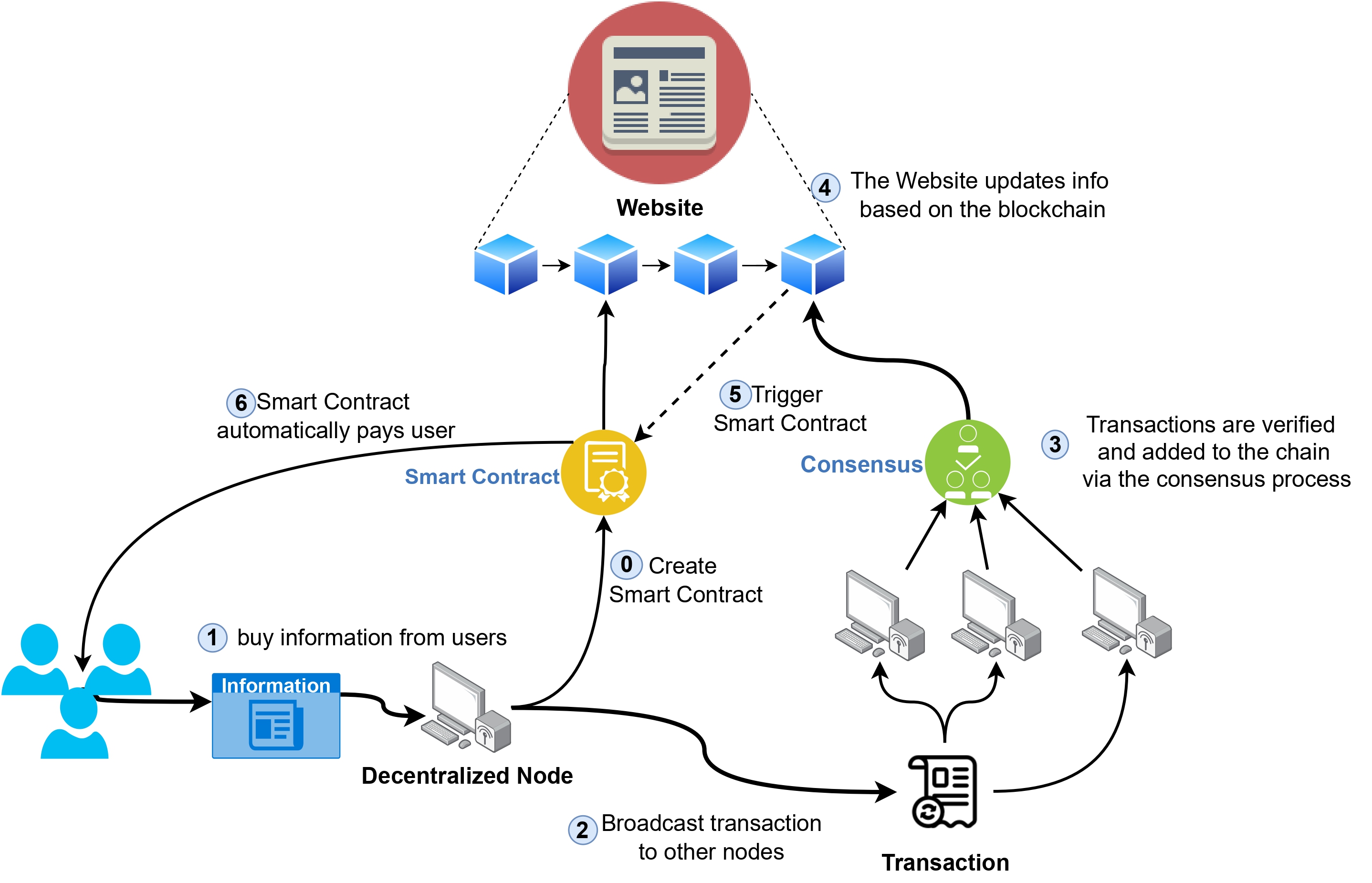}
	\caption{Illustrations of our proposed blockchain-based information management for Web 3.0.}
	\label{design_model}
\vspace{-20pt}
\end{figure}

	As demonstrated in Fig.~\ref{design_model}, there are some key components in our proposed SBW framework as follows: 
\begin{itemize}
 \item \emph{Website}: In our system, a website serves as a repository for information that users may access, including data collected in a decentralized manner.
\item \emph{Decentralized mining nodes}: are core components of blockchain systems, assuring the blockchain's decentralization and security. In addition, in our proposed SBW framework, the nodes can buy information directly from the users to contribute to the website via the consensus process. Nodes selected to be the leader in the consensus process can obtain rewards (e.g., in the form of network tokens) for their contributions.
\item \emph{Users}: are the main parties to collect and provide information to the system. Specifically, the users can gather information and sell it to the nodes for benefits (e.g., they can receive payments in the form of network tokens for the information they sell to the nodes).
 \item \emph{Blockchain}: The blockchain stores and manages the information of the website, as well as the digital assets (i.e., network tokens) of the nodes and users.
 \item \emph{Smart contracts}: Smart contracts are user-defined programs that will be automatically triggered when the conditions within are met. Our system creates smart contracts to facilitate the information buying and selling processes between the users and the nodes.
 \item \emph{Consensus mechanism}: is the core of the blockchain, governing all blockchain processes and ensuring that data stored in the blockchain cannot be modified without the majority of users and nodes consenting. In our system, the consensus mechanism incentivizes users and nodes to contribute more information to the website.
 \end{itemize} 

The data management process of the system is as follows:
\begin{itemize}
        \item \textit{Step 0:} A smart contract is created between a node and a user. This smart contract specifies the payment for each unit of information that the node buys from the user.
	\item \textit{Step 1:} The node chooses the amount of information to buy from the users. 
	\item \textit{Step 2:} The node then broadcasts transactions containing the information (bought from the users) and the payment for the purchased information.
	\item \textit{Step 3:} If a transaction is valid, it will be verified and added to the chain via the consensus process.
	\item \textit{Step 4:} When that transaction has been successfully verified, the website updates information based on transactions stored in the blockchain.
	\item \textit{Step 5:} Once the transaction is stored in the blockchain, it also prompts the smart contract to pay for the user who contributes the information.
	\item \textit{Step 6:} The user will receive the payment from the smart contract regarding their contributed information.
\end{itemize}

\subsection{Proposed Consensus Mechanism}
The consensus mechanism is a crucial element of a blockchain network since it guarantees everyone acknowledges the network's stability under trustless settings and controls other network functions such as adding new transactions and motivating users to behave appropriately~\cite{nguyen2019proof}. The majority of current blockchain networks utilize the PoW consensus mechanism. However, PoW has various drawbacks, including high energy consumption and noticeable latency~\cite{digiconomist}. For example, the Bitcoin network can only handle seven transactions per second and might take up to an hour to finalize a transaction~\cite{nguyen2019proof}. These limitations significantly hinder blockchain integration in the development of Web 3.0, which requires control of a vast amount of data in a time-sensitive manner. Recently, a new consensus process, called PoS, was created to address these issues. PoS techniques were developed as energy-saving replacements for PoW. In PoS, leaders are selected depending on their stakes in the blockchain network rather than their processing power resources. In particular, the stakes of a node in the PoS consensus process are defined by the number of digital tokens, such as coins in cryptocurrencies, it stores. As a result, PoS has significant benefits over PoW, such as low energy consumption, faster transaction confirmation time, and higher transaction throughput. These advantages make the PoS consensus mechanism more suitable for Web 3.0 applications.

Therefore, this paper proposes a PoS-based consensus mechanism for our framework. Unlike the conventional PoS mechanism, our proposed consensus mechanism determines the stakeholder's stake based on their contributions to the network. As a result, the nodes with higher information contribution have a higher opportunity to be selected as the leader and get the block rewards. Notably, the possibility $\textrm{Pr}_n$ that node $n$ is chosen to be the network's leader to get rewards over $N$ participants is: 
 \begin{equation}
	\textrm{Pr}_n=\dfrac{S_n} {\sum_{i=1}^N S_i},
 \label{eq:leader_sel_prob}
\end{equation} 
where $S_n$ represents the information contribution of node $n$. This means that the more information a node contributes to the system, the more likely it will be chosen as the leader. As observed from equation (1), nodes in our system can contribute information to earn rewards similar to the amount of information they contribute. However, if a node increases its contribution, the other nodes' rewards will be decreased. This is a conflict of interest. Thus, in the next section, we use game theory to analyze how nodes' strategies affect the system and how to incentivize nodes to contribute more to the system.
 \vspace{-2pt}

\section{Game Formulation and Analysis}
\vspace{-4pt}	
In the considered system, there is a set $\mathcal{N}=(1,\ldots, N)$ of Web 3.0 decentralized nodes that will collect information from a set $\mathcal{M}=(1,\ldots, M)$ of users to contribute. Each node $n$ can freely choose the amount of information it wants to buy from each user $m$, denoted by $s_n^m$. Moreover, each pair of nodes/users has a fixed unit cost, denoted by $C^m_n$. The difference in cost is because the connection between a node and a user might affect the communication cost. Meanwhile, the consensus time of each round is limited, and thus, the node cannot wait too long to receive the information from the user. Therefore, the cost reflects both the monetary price (the node pays the user) and the transmission cost. Moreover, each user $m$ only has a maximum amount of information it can sell, denoted by $I_m$. Using all the information it buys from the users, node $n$ participates in the consensus process to have a chance to earn the block reward $R$. Particularly, the utility of node $n$ can be determined by:

\begin{equation}
	\label{nodeuti}
	U_n=\dfrac{\sum_{m=1}^M s_n^m}{\sum_{n=1}^{N} \sum_{m=1}^M s_n^m} R- \sum_{m=1}^M C^m_n s_n^m.
\end{equation} 

As observed from~\eqref{nodeuti}, the utility of a node increases as the amount of information it buys from the user increases. However, it also decreases if the other nodes buy more user information. Therefore, this conflict of interest can be assessed using the non-cooperative game theory~\cite{han2012game}. 

Non-cooperative game is one of the most significant sub-fields of game theory, which models scenarios with conflict of interest among participants/players. Each participant in a non-cooperative game acts rationally to maximize its profit, which is influenced by the actions of every other player. Formally, the game can be denoted by $\mathcal{G}(\mathcal{N},(\mathcal{S}_n)_{n\in\mathcal{N}}$,$(U_n)_{n\in \mathcal{N}})$, where $\mathcal{N}$ is the set of players (mining nodes in our blockchain system). The strategy set $\mathcal{S}_n$ contains all methods that might be used by player $n$, and the reward (utility) function for each player is denoted by $U_n$. Particularly, the strategy space of the players can be defined by:

\begin{equation}
\begin{aligned}
   \mathcal{S}_n = \{\mathbf{s}_n = [{s^1_n},\ldots,{s^M_n}] |\sum_{n=1}^N s^m_n \leq I_m\}. 
   \end{aligned}
\end{equation}

In non-cooperative games, every player acts rationally to increase utility while considering the other players' strategies. Formally, for any given set of strategies of all the other players, player $n$ will choose its strategy according to the following best response function:

\begin{equation}
\mathbf{b}_n (\mathbfcal{S}_{-n}) \leftarrow \{\mathbf{s}_n \in \mathcal{S}_n|U_n(\mathbf{s}_n)\geq U_n(\mathbf{s}'_n),\forall \mathbf{s}'_n \in S_n\},
\end{equation}

where $\mathbf{b}_n$ denotes the best response of player $n$, $\mathbf{s}_n$ and $\mathbf{s}'_n$ denote two arbitrary strategies of player $n$, and $\mathbfcal{S}_{-n}$ denotes a fixed strategies set of all players except player $n$. When all the other players choose $\mathbfcal{S}_{-n}$, $\mathbf{b}_n$ will maximize player $n$'s utility. The pure strategy Nash equilibrium is a good result of a non-cooperative game \cite{han2012game}. At the Nash equilibrium, no player can unilaterally modify its strategy to improve its utility. As a result, the system becomes stable at the Nash equilibrium because rational players will not change their strategies. In the following theorem, we first prove that the Nash equilibrium exists in $\mathcal{G}$.
\begin{theorem}
At least one Nash equilibrium exists in $\mathcal{G}$.
\end{theorem}
\begin{IEEEproof}
According to \cite{han2012game}, if $S_n$ is compact and convex and $U_n$ is quasi-concave $\forall n\in\mathcal{N}$, there exists at least one Nash equilibrium in this game. We thus first prove that the utility function of every player is concave over its possible strategy sets. The partial derivative of second order of $U_n$ with respect to $s_n^m$ is:
\begin{equation}
	\label{eq:diff1}
	\begin{split}
		\dfrac{\partial^2 U_n}{\partial (s_n^m)^2}=\dfrac{-2RT}{(\sum_{n=1}^{N} \sum_{m=1}^M s_n^m)^3},
	\end{split}
\end{equation}

where $T$=$\sum_{i \in \mathcal{N}_{-n}} \sum_{m=1}^M s^i_m$ expresses the total information of all players except player $n$. Since $R>0$, $T>0$, and $(\sum_{n=1}^{N} \sum_{m=1}^M s_n^m) >0$, we have  $\dfrac{\partial^2 U_n}{\partial (s_n^m)^2} < 0, \forall s_n^m$. As a result, $U_n$ is concave over $\mathbf{S}_n$. Additionally, $\mathbf{S}_n$ is defined as a compact and convex set for every player. As a result, there exists at least one Nash equilibrium in this game
\end{IEEEproof}

With the Nash equilibrium's existence established, we implement an iterative algorithm (Algorithm 1) to find the Nash equilibrium. In Algorithm 1, the strategies of all players are first randomly initialized. Then, we fix the strategies of all players except player 1 and proceed to find the best response of player 1 by maximizing its utility function. After the best response for player-1 is found, we set it as the new strategy of player-1 and continue to find the best response for player 2 similarly. This loop continues to find the best responses for all players, and the loop is repeated until no player changes its strategy for the entire loop. Once this termination condition is met, all the best responses constitute the Nash equilibrium. 

\begin{algorithm}
\caption{Iterative algorithm to find Nash equilibrium in $\mathcal{G}$}\label{alg:cap}
\begin{algorithmic}[1]
\State Initialize $\mathbf{s}_n$ \Comment{Randomly initialize all strategies}
\Repeat
\For{$n=1:N$} 
\State $\mathbf{b}_n \leftarrow {\mathbf{s}_n \in \mathcal{S}_n|U_n(\mathbf{s}_n)\geq U_n(\mathbf{s}'_n),\forall \mathbf{s}'_n \in S_i}$ \LineComment{Find best response of player n}
\State $\mathbf{s}_n \leftarrow \mathbf{b}_n$ \Comment{Fix player n strategy}
\EndFor
\Until No player changes strategy
\end{algorithmic}
\end{algorithm}

\vspace{-2pt}

\section{Numerical Results}
\label{sec:simu}

In this section, we conduct comprehensive simulations to assess the impact of individual player strategies on the system and demonstrate the convergence of the proposed algorithm. To begin, we explore the structure of the user's utility function in a system with 4 decentralized nodes and 6 users. Subsequently, we demonstrate the convergence of nodes' strategies to equilibrium when employing the proposed Proof-of-Stake (PoS)-based consensus mechanism. We further explore the effects of changing the unit cost of a node and the block reward on user strategies and utilities. Finally, we expand our analysis to accommodate up to 75 nodes and 100 users, which closely resembles a real-world deployment scenario. 

In the first experiment, we simulate a system with $N=4$ and $M=6$. We also set $R=1000$ based on a real-world PoS-based blockchain network~\cite{cardano2019digital}. The total amount of data a user can sell is limited to 150 units. The unit cost of purchasing data for each node/user pair is randomly generated in the range $[1.0, 2.0]$ to reflect the heterogeneity in network transmission cost. Here, we check the utility function of a single node and determine the optimal strategy for that node when the strategies of the other nodes are fixed. The initial strategies of all players are to buy 2 units of data from every user, i.e., $s_n^m = 2, \forall n=1,\ldots,4$, and $m=1,\ldots,6$. %Here, we use SciPy's optimization framework to maximize the utility function and find the optimal strategy for a node in each iteration of Algorithm~\ref{alg:cap}, when the strategies of other nodes are fixed.% Algorithm~\ref{alg:cap} is also used to simulate the non-cooperative game and record the utilities and strategies of all participating nodes after each iteration to demonstrate the convergence of the node strategies to the Nash equilibrium. 

 \begin{figure}[!]
\centering
    \includegraphics[width=0.4\textwidth]{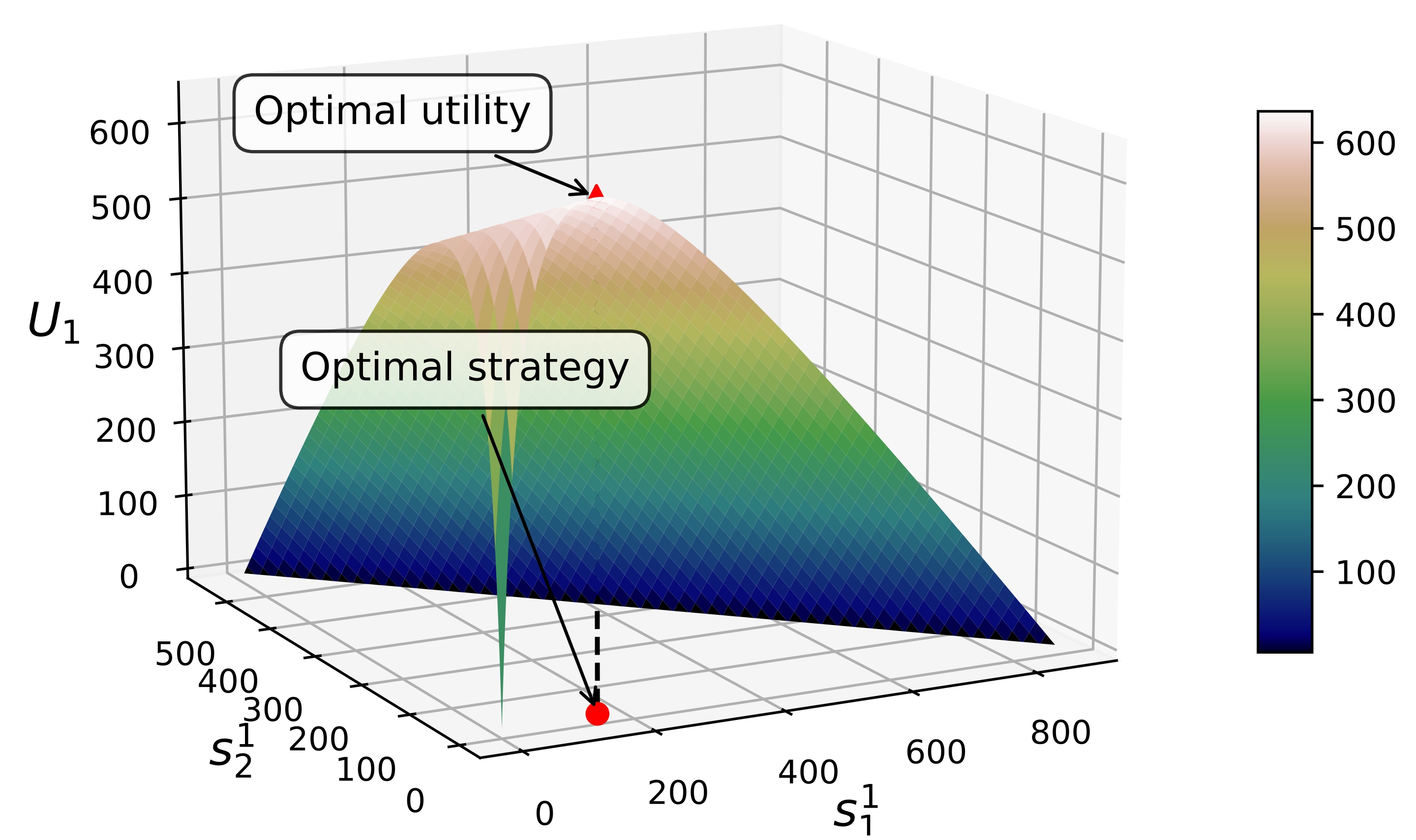}
    \caption{Illustration of the utility function.}
    \label{fig:utility}
     \vspace{-17pt}
\end{figure}

Fig.~\ref{fig:utility} demonstrates the utility function and the optimal strategy of Node 1 at the beginning iteration when the strategies of the other nodes (i.e., Nodes 2, 3, and 4) are fixed. Here, we plot the utility of Node 1 in 3D space as a function of $s_1^1$ and $s_1^2$, while the other $s_1^m, \forall m=3,4,5,6$ are fixed at 2, i.e., the initial values. This figure shows that Node 1 can achieve the optimal utility ($U_1^* = 640.6$) when it buys $s_1^1 = 144$ units of data from User 1 and $s_1^2 = 0$ units from User 2. It is noted that Node 1 does not choose to purchase all of the available data from User 1 and User 2, even though the remaining amounts of data from these users are 144. The reason is that there is a certain threshold where the cost of buying more data exceeds the additional reward gained. This threshold depends on the unit costs of purchasing data at node-1 and the strategies used by the other nodes in the system. 

\begin{figure}[!]
     \centering
     \begin{subfigure}[b]{0.31\textwidth}
         \centering
         \includegraphics[width=\textwidth]{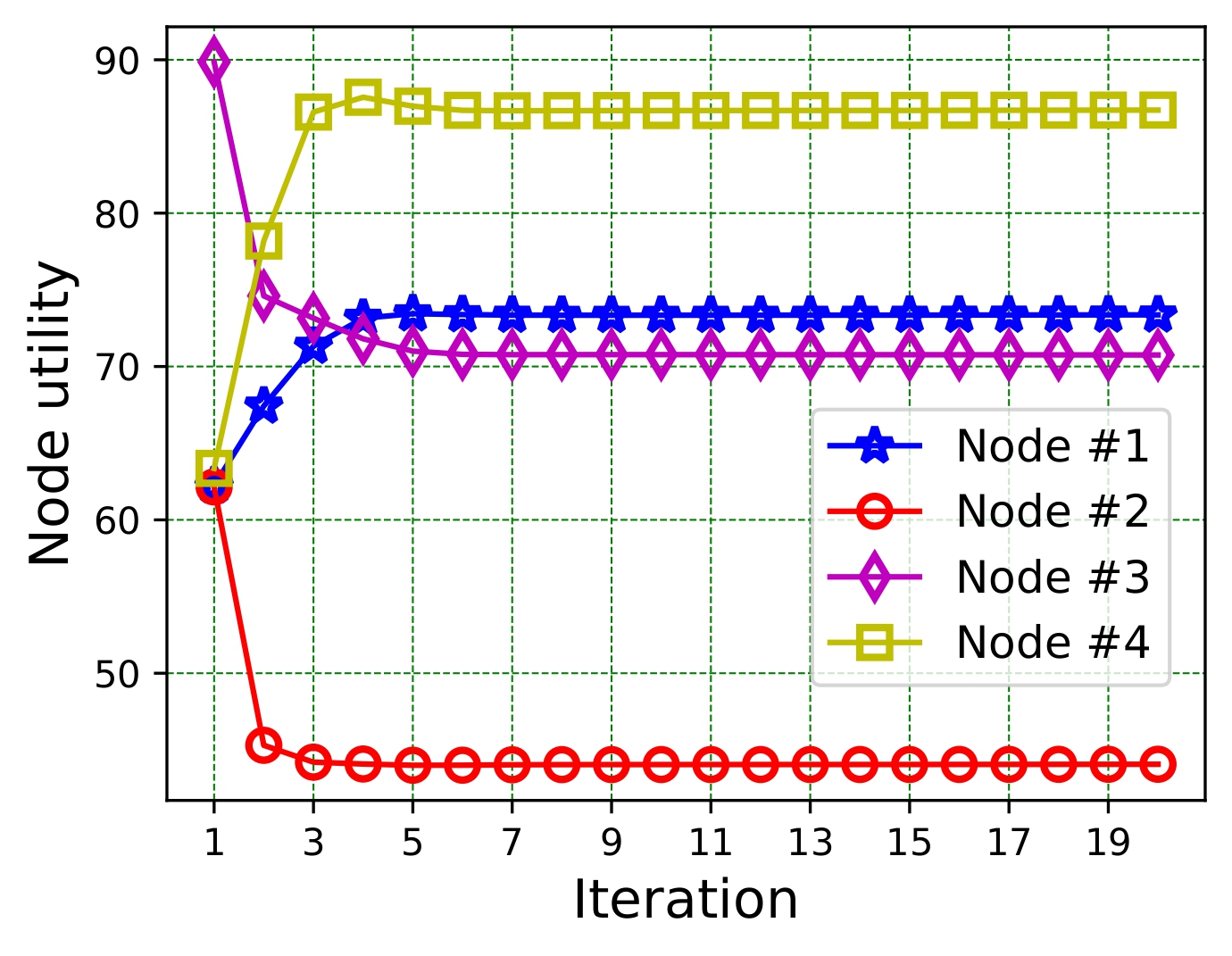}
         \caption{Default setting.}
         \label{fig:convergence_1}
     \end{subfigure}
     \hfill
     \begin{subfigure}[b]{0.31\textwidth}
         \centering
         \includegraphics[width=\textwidth]{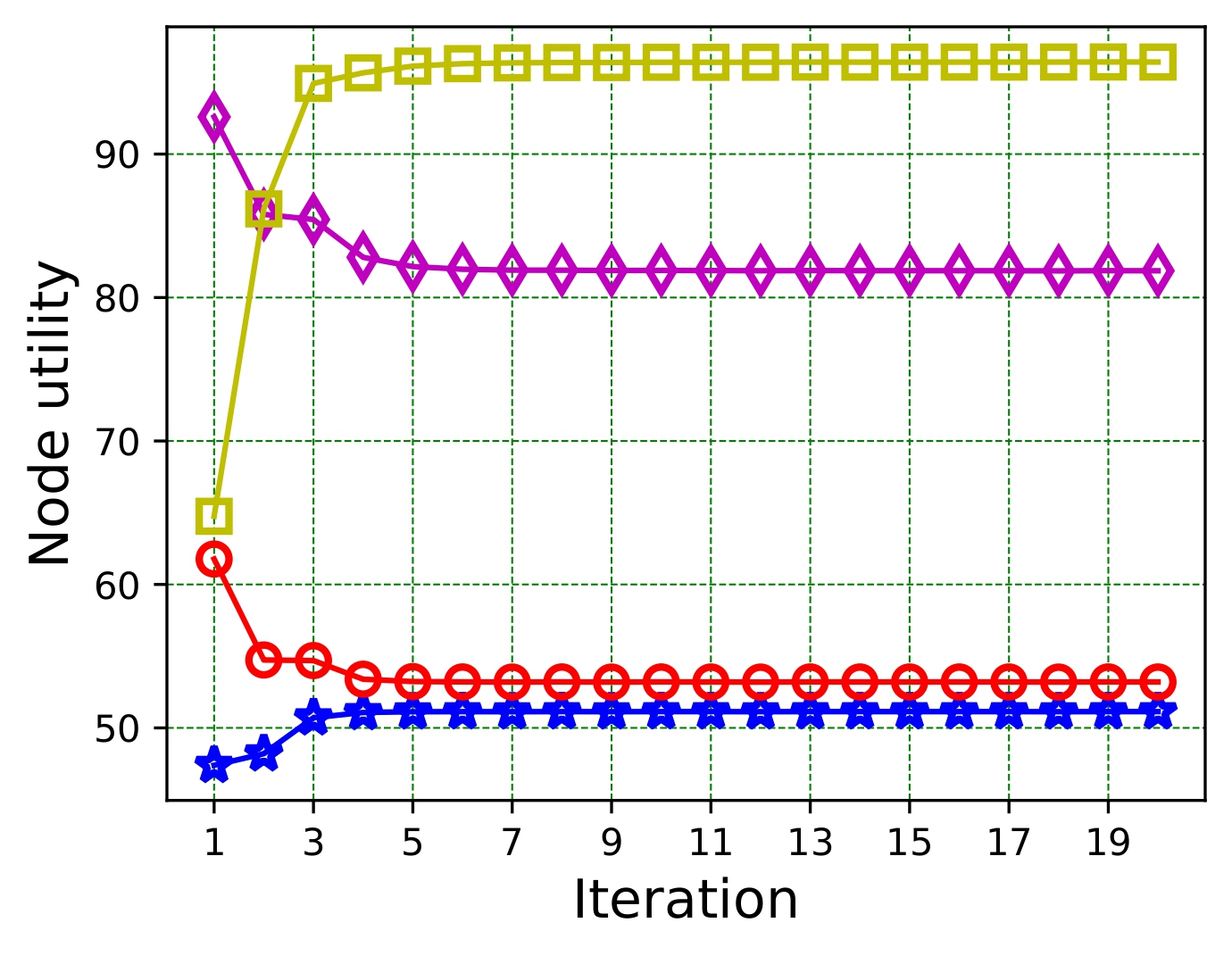}
         \caption{Increasing unit costs of node 1 by 10\%}
         \label{fig:convergence_2}
     \end{subfigure}
     \hfill
     \begin{subfigure}[b]{0.31\textwidth}
         \centering
         \includegraphics[width=\textwidth]{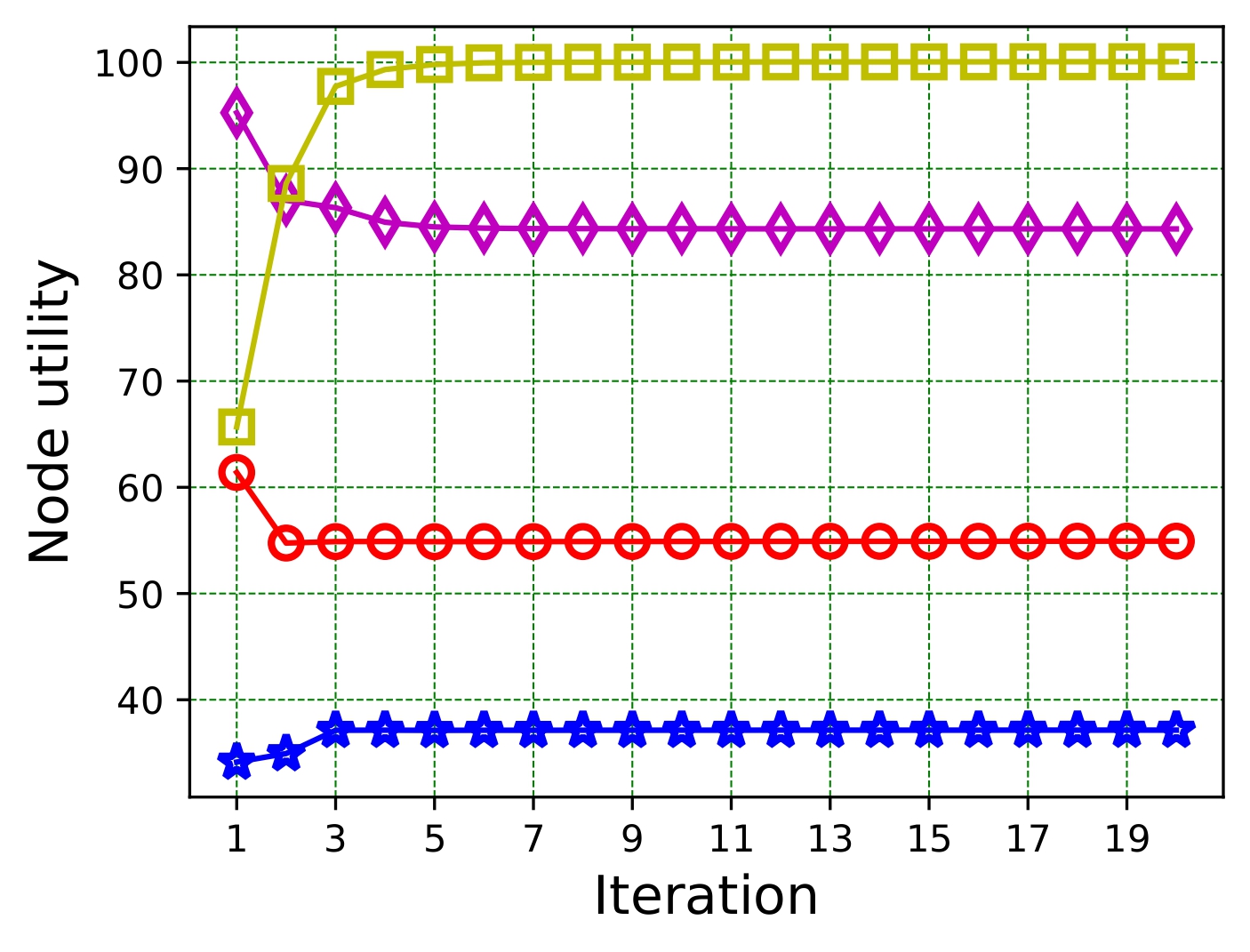}
         \caption{Increasing unit costs of node 1 by 20\%}
         \label{fig:convergence_3}
     \end{subfigure}
        \caption{Convergence of the iterative algorithm.}
        \label{fig:convergence}
        \vspace{-24pt}
\end{figure}

Fig.~\ref{fig:convergence}\subref{fig:convergence_1} shows the convergence of the nodes' strategies to the equilibrium. According to the figure, this convergence occurs after 7 iterations. At the equilibrium, Nodes 1 and 4 have the highest utilities because they are offered lower costs by the users, resulting in a better cost-to-reward ratio. In contrast, Node 2 has the lowest utility because it has the highest purchasing cost. %, i.e., more than 1.1 per data unit. 
% Specifically, node-1 only purchases data from users 1 and 6, where it can obtain the lowest cost per unit (i.e. $C_1^1 = C_1^6 = 1.0$), while node 2 buys data from users 4, 5, and 6.
Next, we investigate how unit cost affects the convergence to the equilibrium and the players' utilities at that point. To do this, we increase the unit cost of Node 1 by 10\% and 20\% from every user and plot the convergence curves of the players' utilities in Fig.~\ref{fig:convergence}\subref{fig:convergence_2} and Fig.~\ref{fig:convergence}\subref{fig:convergence_3}. As shown in the figures, all the players' strategies converge after 7 iterations. More notably, when the unit cost of Node 1 increases by 10\% and 20\%, the total amounts of purchased data at Node 1 are reduced by 24.74\% and 31.27\%, respectively, since Node 1 is less interested in buying data due to a higher cost-to-reward ratio. As a result, its utility at the equilibrium is reduced by 30.3\% and 49.39\%, respectively.  In contrast, the utilities of all other nodes increase as they have a higher chance to earn the block reward, as reflected in~(\ref{eq:leader_sel_prob}).
%Also, as the cost of purchasing data at node 1 grows, node 1 has less conflict with node 4 in buying data from user 6 (note that in the default setting, both nodes can purchase data from user 6 with the unit cost of 1.0). Thus, its strategy and utility converge faster, i.e., after 5 iterations and 3 iterations as shown in Fig.~\ref{fig:convergence}\subref{fig:convergence_2} and Fig.~\ref{fig:convergence}\subref{fig:convergence_3}, respectively.
  
\begin{figure}[!]
    \includegraphics[width=0.46\textwidth]{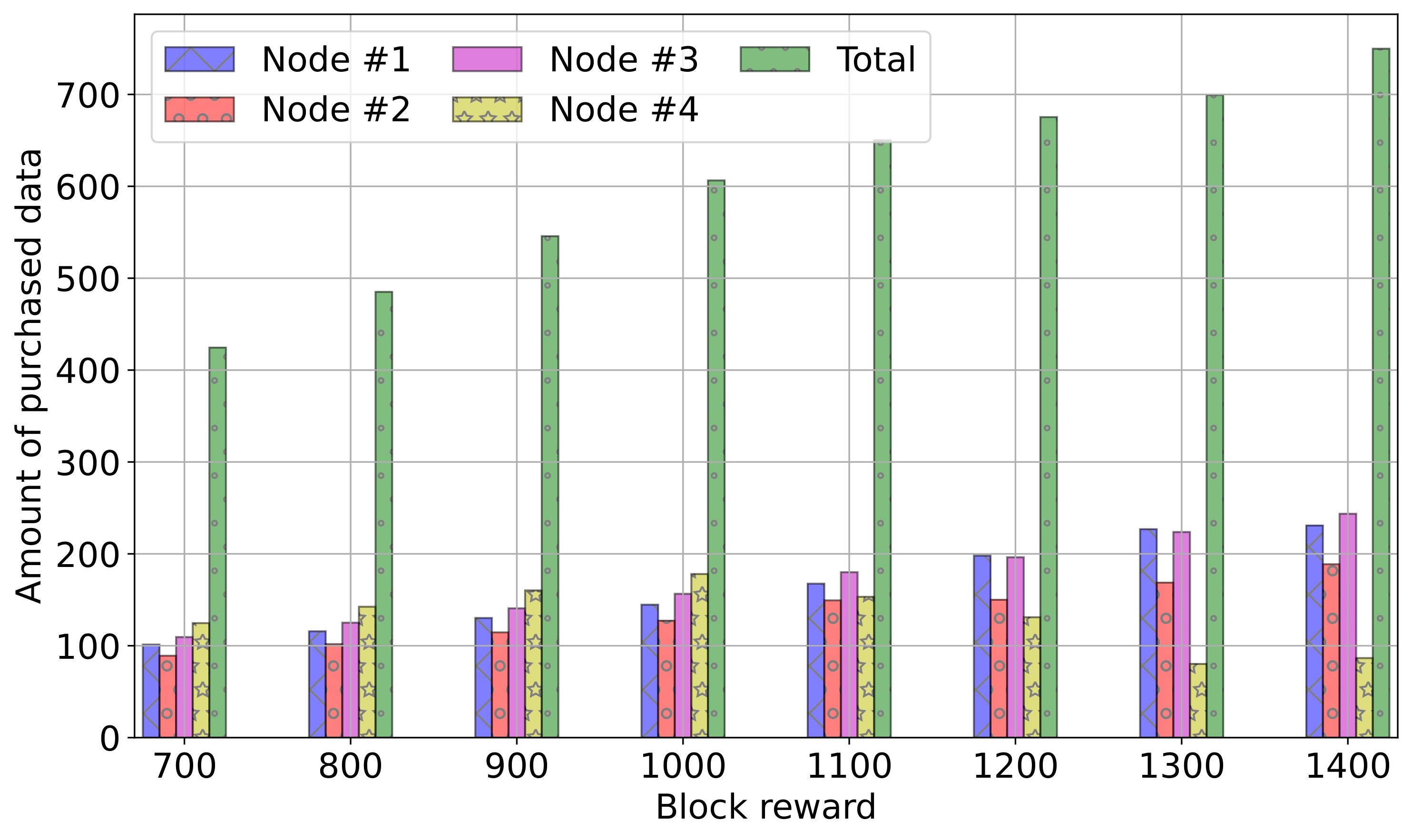}
    \caption{Amount of purchased data by all nodes when varying the block reward $R$.}
    \label{fig:vary_R}
    \vspace{-10pt}
\end{figure}

Subsequently, Fig.~\ref{fig:vary_R} demonstrates how the amount of data each node purchases changes when the block reward varies between 700 and 1400. The figure shows that the total amount of data purchased in the system increases linearly with the block reward $R$. The reason is that a lower cost-to-reward ratio gives the nodes greater appeal to buy data from the users. However, we do not notice the same linear increasing trend in the total amount of purchased data at each node individually since one's strategy depends on the others. For example, when $R = 1000$, Node 4 can buy $s_4^5 = 145.04$ units of data from User 5 because the other nodes have less interest in purchasing data from this user. When the value of $R$ is increased to 1300, the cost-to-reward ratio for every node becomes lower. In response, Nodes 2, 3, and 4 compete to purchase data from User 5 to gain more rewards. At the equilibrium point, Node 4 buys $s_4^5 = 78.51$ units of data from User 5, while Node 2 and Node 3 buy $s_2^5 = 20.68$ and $s_3^5 = 53.16$ units of data, respectively. This non-cooperative nature of the system leads to a decrease in the total amount of purchased data at Node 4, as shown in Fig.~\ref{fig:vary_R}.
%where they can find the lowest unit cost 

 \begin{figure}[!]
\centering
     \includegraphics[width=0.4\textwidth]{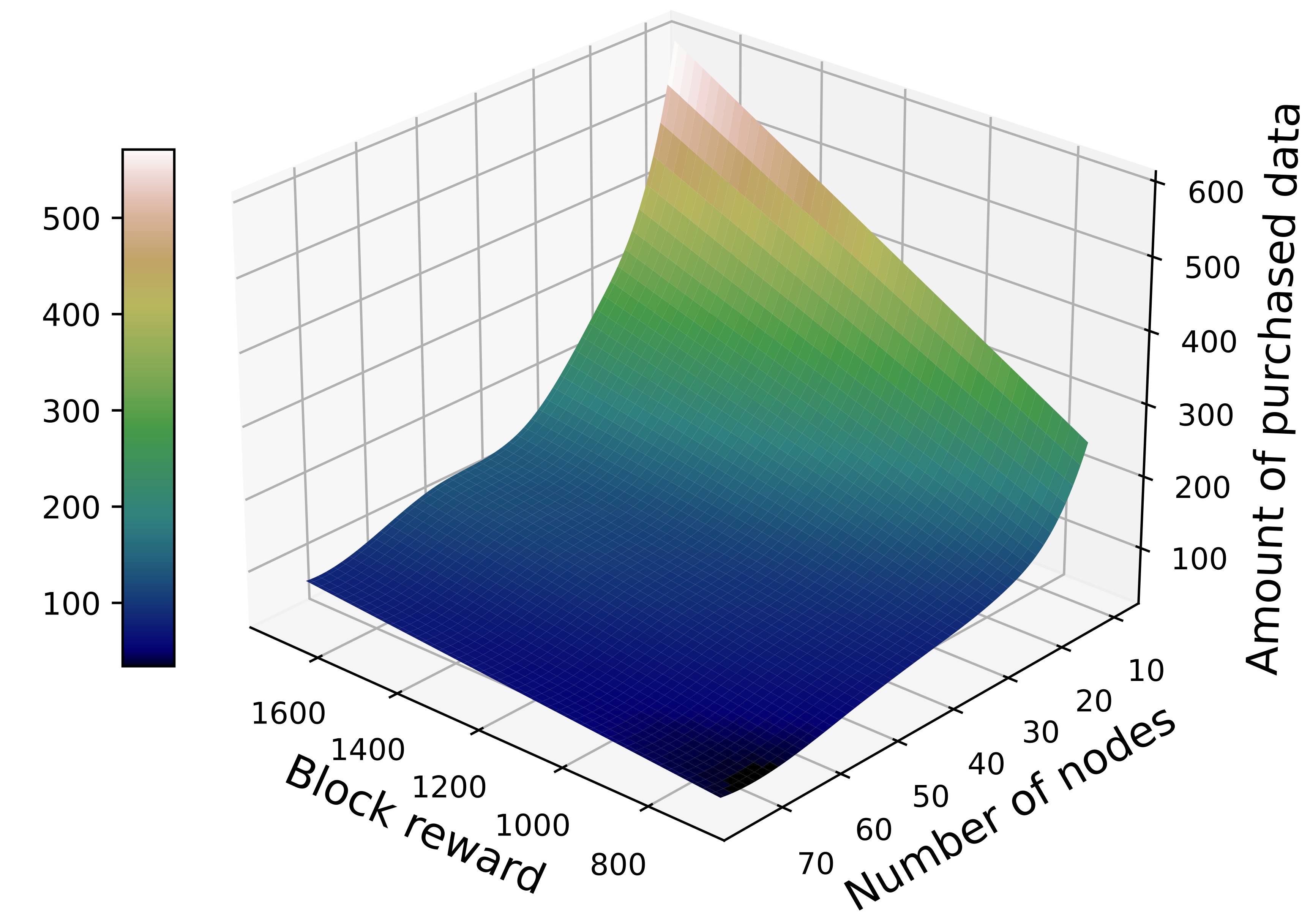}
    \caption{Impact of varying the block reward and number of nodes on the total amount of data purchased in the system.}
    \label{fig:data_purchased}
     \vspace{-10pt}
\end{figure}

Finally, Fig.~\ref{fig:data_purchased} demonstrates the impact of varying the block reward and the number of nodes on the total amount of data purchased in the system. Here, the block reward ranges from 700 to 1700, while the number of nodes varies from 10 to 75. A clear linear correlation emerges as the block reward increases, indicating that higher rewards incentivize nodes to procure more user data. This behavior is due to the improved cost-to-reward ratio for the nodes, making data purchases more attractive as the potential rewards increase. However, an increase in the number of nodes leads to a reduction in the total data acquired. This reduction is due to increased node competition for purchasing user data, resulting in nodes adopting more conservative strategies, such as purchasing less data, to manage costs and maximize utility. As a result, each node may purchase fewer data units, leading to a lower total data purchased across the system. 
	
\section{Conclusion}
	\label{sec:Sum}
In this paper, we have proposed a novel blockchain-based information management framework for Web 3.0, namely SBW, which utilizes smart contracts to automate complex information management processes. Moreover, we have developed a new consensus mechanism to incentivize users and blockchain nodes to contribute information to the website. Furthermore, we have formulated and analyzed the behaviors of nodes and users behaviors in our system using non-cooperative game theory. Additionally, we have conducted various numerical experiments to verify our analysis and study essential parameters in the proposed system. The findings confirm our theoretical analysis and demonstrate that our proposed consensus mechanism can increase user profits and incentivize users to contribute more to the system.

\bibliographystyle{IEEE}

\end{document}